\documentclass[twocolumn]{aastex6}
\usepackage{amsmath}

\fullcollaborationName{}

\begin{document}


\title{ The  {\em Chandra} COSMOS legacy survey:   Energy Spectrum of the Cosmic X-ray Background and constraints on  undetected populations }


\author{Nico Cappelluti\altaffilmark{1,2,3}, Yanxia Li\altaffilmark{4,1,2,3}, Angelo Ricarte\altaffilmark{3}, Bhaskar Agarwal\altaffilmark{1,2,3}, Viola Allevato\altaffilmark{5}, Tonima Tasnim Ananna\altaffilmark{1,2,3}, Marco Ajello\altaffilmark{6}, Francesca Civano\altaffilmark{7}, Andrea Comastri\altaffilmark{8}, Martin Elvis\altaffilmark{7},  Alexis Finoguenov\altaffilmark{5,9}, Roberto Gilli\altaffilmark{8}, G\"unther Hasinger\altaffilmark{4}, Stefano Marchesi\altaffilmark{6}, Priyamvada Natarajan\altaffilmark{1,2,3}, Fabio Pacucci\altaffilmark{2},   E. Treister\altaffilmark{10} \and C. Megan Urry\altaffilmark{1,2,3},
 }

\and


\altaffiltext{1}{Yale Center for Astronomy and Astrophysics, P.O. Box 208121, New Haven, CT 06520.}
\altaffiltext{2}{Department of Physics, Yale University, P.O. Box 208121, New Haven, CT 06520.}
\altaffiltext{3}{Department of Astronomy, Yale University, PO Box 208101, New Haven, CT 06520.}
\altaffiltext{4}{Institute for Astronomy, 2680 Woodlawn Drive, University of Hawaii, Honolulu, HI 96822, USA}
\altaffiltext{5}{Department of Physics, University of Helsinki, Gustaf H\"allstr\"omin katu 2a, FI-00014 Helsinki, Finland}
\altaffiltext{6}{Department of Physics and Astronomy, Clemson University, Kinard Lab of Physics, Clemson, SC 29634-0978, USA}
\altaffiltext{7}{Harvard-Smithsonian Center for Astrophysics, 60 Garden Street, Cambridge, MA 02138, USA}
\altaffiltext{8}{INAF-Osservatorio Astronomico di Bologna, via Ranzani 1, I-40127 Bologna, Italy}
\altaffiltext{9}{Max-Planck-Institut f\"ur extraterrestrische Physik, Postfach 1312, 85741, Garching bei MŸnchen, Germany}
\altaffiltext{10}{Instituto de Astrofisica, Facultad de Fisica, Pontificia Universidad Catolica de Chile, Casilla 306, Santiago 22, Chile}
\begin{abstract}
Using   {\em Chandra}  observations in the 2.15 deg$^{2}$ COSMOS legacy field, we present one of the most accurate measurements of the Cosmic X-ray Background (CXB) spectrum to date in the [0.3-7] keV energy band.  The CXB has three distinct components: contributions from two Galactic collisional thermal plasmas at kT$\sim$0.27 and 0.07 keV and an extragalactic  power-law with photon spectral index $\Gamma$=1.45$\pm{0.02}$. The 1 keV normalization of the extragalactic component is 10.91$\pm{0.16}$  keV cm$^{-2}$ s$^{-1}$ sr$^{-1}$ keV$^{-1}$. 
Removing all  X-ray detected sources, the remaining unresolved CXB is best-fit by a power-law with normalization 4.18$\pm{0.26}$ keV cm$^{-2}$ s$^{-1}$  sr$^{-1}$ keV$^{-1}$ and photon spectral index $\Gamma$=1.57$\pm{0.10}$. Removing faint galaxies down to i$_{AB}\sim$27-28 leaves a hard spectrum with $\Gamma\sim$1.25 and a
1 keV normalization of $\sim$1.37 keV cm$^{-2}$ s$^{-1}$ sr$^{-1}$ keV$^{-1}$.
This means that $\sim$91\% of the observed CXB is resolved into detected X-ray sources and undetected  galaxies. 
 Unresolved sources that  contribute $\sim 8-9\%$ of the total CXB show a marginal evidence of being harder and possibly more obscured than resolved sources. Another  $\sim$1\% of the CXB can be attributed to still undetected star forming galaxies and  absorbed AGN. 
 According to these limits, we investigate a scenario where  early black holes  totally account for non source CXB fraction and constrain some of their properties. In order to not exceed the remaining CXB and the  $z\sim$6 accreted mass density, such a population of black holes must grow in Compton-thick  envelopes with N$_{H}>$1.6$\times$10$^{25}$ cm$^{-2}$ and form in extremely low metallicity environments $(Z_\odot)\sim10^{-3}$.  

\end{abstract}




\section{Introduction} \label{sec:intro}

Focusing X-ray telescopes like ROSAT,    {\em Chandra}, XMM-{\em Newton} and {\it Swift},
 have  shown that the main contributors to the extragalactic Cosmic X-ray Background (CXB) are Active Galactic Nuclei (AGN). Although the spectrum of the CXB has been measured by almost every X-ray telescope,  measurements vary significantly in the [0.3-10] keV energy band. The actual normalization of the CXB spectrum is therefore still a matter of debate and this uncertainty leaves systematic uncertainties in  AGN population synthesis models \citep{gilli01,treist05,gilli07,treist09}. An important tool to fully understand the nature of the CXB is the unresolved 
CXB spectrum, once faint X-ray and optical/NIR sources have been removed. Thanks to its excellent angular resolution, 
Chandra can resolve faint sources in deep exposures, which can then be excised. This allows us to study both the X-ray stacked spectrum and the remaining CXB flux.

Previous mission measurements,  agree that the CXB spectral index is $\Gamma\sim$1.4 
but the normalization is uncertain by $\sim$20-30\%.
These discrepancies are likely due to
inaccurate spectral cross-calibrations, poorly understood  instrumental backgrounds, and cosmic variance  \citep{more09}.  Another significant limitation in determining the amplitude of the soft extragalactic CXB is the ability to remove contamination from galactic components that peak below 2 keV where the effective area of focusing X-ray telescopes peaks.
This poses a serious challenge to the understanding of the true fraction of unresolved soft CXB. 

Recent papers suggest that the unresolved CXB may contain important information on the first generation of massive black holes in the Universe \citep{salva,cap12,cap13}. While deep surveys provide an estimate of the fraction of unresolved CXB, via the integration of number counts \citep{more03,wor04,hm07,more12}, the spectrum of the unresolved background has never been measured with sufficiently deep and wide surveys.    \citet{hm06,hm07} successfully removed local foregrounds to make this measurement in the   {\em Chandra}  deep fields, but the limited size of the fields meant it was cosmic variance limited at the 20-30\% level. This impels us to carefully study the unresolved CXB with the highest degree of accuracy currently permitted by data. 

In this paper, we make use of the best available dataset, which is the largest deep survey ever performed  by  {\em Chandra}:  the COSMOS-Legacy survey \citep{elvis,civ16,mar16}. Here, we  present a novel precise measurement of the CXB,  its unresolved fraction, and new constraints on the properties of z$>$6 black holes by using the Soltan argument.

\section{Dataset and analysis}

The    {\em Chandra}        COSMOS-Legacy survey \citep[CCLS,][]{elvis,civ16}  is  an X-ray Visionary Program that imaged the 2.2 deg$^2$ 
COSMOS field  \citep{scov} for a total of 4.6 Ms. The survey has an effective exposure of 160 ks over the central 1.5 deg$^2$ and 
of $\sim$80 ks elsewhere. A total of 4016 X-ray sources are detected down to 
flux limits of 2.2$\times$10$^{-16}$, 1.5$\times$10$^{-15}$, and 8.9$\times$10$^{-16}$ erg cm$^{-2}$ s$^{-1}$ in the [0.5--2], [2--10], and [0.5--10] keV  energy bands, respectively. 

All the observations were performed in  VFAINT telemetry mode since it allows a lower instrumental background 
value. 
Here we briefly summarize our analysis, the details of which were mostly reported by \citet{civ16} \citet{pucc09}.
Level 1 data products were processed with the CIAO-tool  $chandra\_repro$  retaining only valid event grades. Astrometry in each pointing was matched with the optical catalogs  of  \citet{cap07} and \citet{ilb09}. Particle background flares were removed using the $deflare$ tool in compliance with the  ACIS background analysis requirement after excising from the dataset. 
In order to minimize uncertainties in modeling the particle quiescent background,
we took special precautions to ensure that background flares were removed in a such a way that residuals from undetected faint flares were reduced.
As shown in \citet{hm06}, the [2.3-7] keV energy range is the most sensitive to particle background flares, and  stowed-mode
observations demonstrate that the [2.3-7] keV to [9.5-12] keV Hardness Ratio (HR) is constant.  They also show that filtering the data for flares only in the [2.3-7] keV energy band results in  missed periods of time during which the background has an anomalous HR. 
To account for this effect, we searched for flares not only in the [2.3-7] keV energy band, but also in the [9.5-12] and [0.3-3] keV bands.  These ``flared'' time intervals were removed from the data \citep{cap09}.  With this procedure, we are confident that the remaining level of flaring is below 1-2\% \citep{hm06}, and therefore 
the amplitude of the quiescent particle background is subject to this level of systematic uncertainty. 
The {\em Chandra} X-ray Center (CXC) ACIS calibration team verified the validity of these assumptions, and no background anomalies had been reported as of May 2016.

X-ray source masking and/or stacking was performed to match the \citet{civ16} X-ray source catalog.  
 For X-ray undetected galaxies, we used the \citet{scova} catalog of $\sim$1 million  detections by the Hubble Space Telescope down to m$_{AB}\sim$27-28 from in $i$-band. This enables a robust removal of faint  X-ray undetected galaxies. Although \citet{ilb09} and \citet{lai} assembled a catalog of $\sim$ 2 million galaxies, using these catalogs would have vastly reduced the area of our spectral extraction (see below) and made comparison with previous works more difficult. Moreover, for these last two catalogs, the coverage and sensitivities are
uneven on the whole survey area. 
\begin{figure}[!th]

\includegraphics[angle=270,scale=0.3]{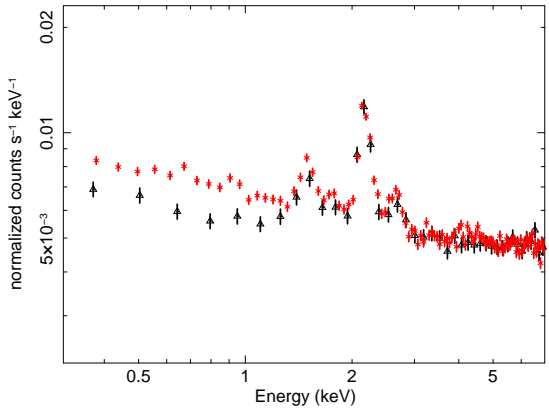}
\caption{ \label{fig:comp}    A comparison of the nsCXB (red stars) before background subtraction and the PIB (black triangles). 
It is worth to note that the error bars of the PIB spectrum are much larger that those of the raw nsCXB spectrum.}
 \end{figure}

\subsection{Spectral extraction}
\begin{figure}[!th]
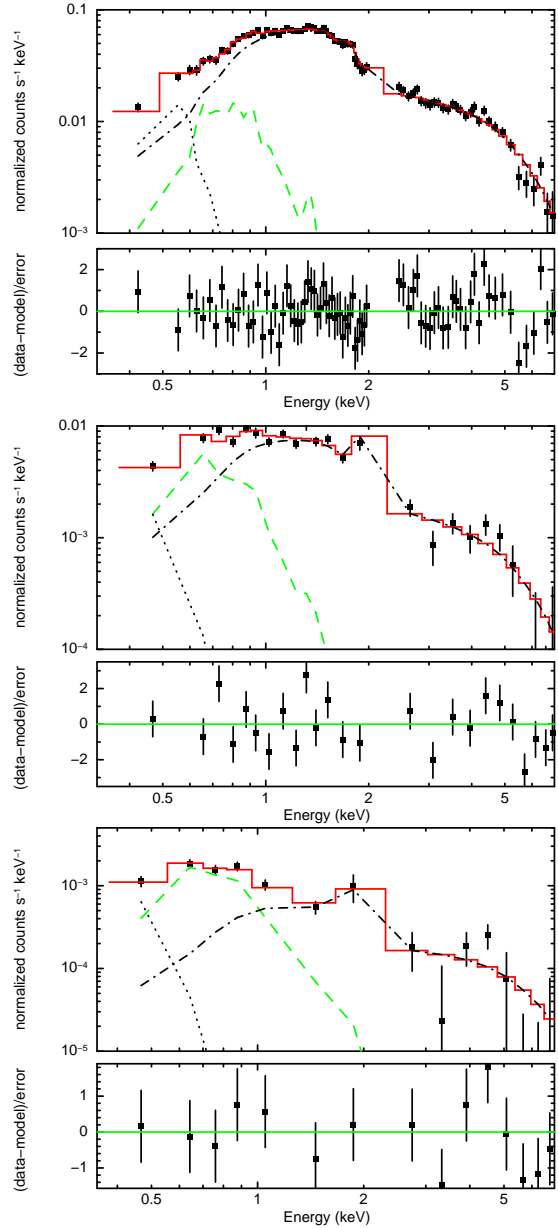

\includegraphics[angle=270,scale=0.3]{full_spec.ps}
\includegraphics[angle=270,scale=0.3]{unres.ps}
\includegraphics[angle=270,scale=0.3]{nonsource_spec.ps}
\caption{ \label{fig:spectra} From top to bottom the binned CXB, uCXB and nsCXB folded spectra, respectively. Black squares are the data points. 
The red continuous line is the best fit model, the dot-dashed-line is the extragalactic components, the green-dashed line is 
the hard thermal component and the dotted line is the local bubble components. In the bottom panels we show the 
fit residuals. Data have been re-binned in order to have at least 10$\sigma$ significance per bin and no more than 20 bins have been combined. 
 }
\end{figure}
To obtain the spectrum of the full CXB, we use the entire ACIS-I area to maximize the collecting and survey area.  This allows us to obtain a measurement which is minimally affected by poor statics or cosmic variance.  We call the spectrum of all photons detected in the field of view (FOV) the CXB spectrum.

However, to extract the {\it unresolved} CXB spectrum, we more carefully select the area used.  Both $Chandra$'s point spread function (PSF) and effective area rapidly degrade with the off-axis angle.  Consequently, at large off axis-angles, little to no usable area is left after excising detected sources.   The {\em Chandra} PSF radius can be approximated with: 
 r$_{90}\sim$ 1$\arcsec$+10$\arcsec(\theta/10\arcmin)^2$, where  r$_{90}$ is 90\% Encircled Energy Radius and $\theta$ is the 
 off-axis angle.  Compromising the need to maximize photon count with the degradation of our observations with off-axis angle, we found that the highest quality data can be obtained using the inner 5$\arcmin$ ($\theta$= 5$\arcmin$ $\implies$ r$_{90} < 3.5\arcsec$).  

To estimate the spectrum of the CXB after removing X-ray sources, we extracted the spectrum of the area remaining in the inner 5$\arcmin$ after removing the sources detected  by \citet{civ16} using a 
7$\arcsec$ radius region around each X-ray centroid. This radius corresponds to twice r$_{90}$, 
which we consider  large enough  to neglect the flux of PSF tails.
Because of the mosaicking, with these choices we still cover most of the CCLS area
and mask 3\% of the pixels because of sources. 
This spectrum will be called  uCXB (unresolved CXB).
  According to Fig. 2 of  \citet{civ16}, and thanks to tiling of the pointings,
this radius safely includes almost the totality of the source fluxes even without limiting the investigation to the inner FOV.
Note that the {\em Chandra} PSF is not circular, but is elongated as a function of the azimuthal angle. Nevertheless, we were able to use circular apertures because the asymmetry of the PSF is washed out by the tiling of the survey, which averages over azimuthal angles. See e.g the treatment of PSF fitting in \citet{cap16}.

To further probe the unknown discrete source CXB  (hereafter nsCXB, non source CXB),
we extracted the spectrum of the area left after removing X-ray sources and HST-ACS detected sources. 
 These sources are so plentiful that using a 3.5$\arcsec$ masking radius leaves little
sky area to perform our measurement. For this reason, we used the approach of \citet{hm07} and limited the  search to the inner 
3.2$\arcmin$ of axis of every pointing. We estimated that r$_{90}<$2.2$\arcsec$, and masked areas around each galaxy of the  \citet{scov} catalog. 
The choice of this radius is a trade-off, ensuring that a large fraction of the optical/NIR selected galaxies X-ray flux is removed and keeping the contamination 
from PSF tails under control (see below for the treatment of PSF tails). 
\\
For each subsample, the net extracted counts  are reported in Table \ref{tab:cts}.
 Remarkably, the CXB spectrum we derive contains  $\sim$123000 net counts.
  For each spectrum, we also computed the field-averaged 
Redistribution Matrix Functions (RMFs) and Ancillary Response Functions (ARF) using the CIAO-tool $specextract$. 
Spectra were then co-added and response matrices averaged after weighting by the exposure time. 

\subsection{Background treatment and systematics}
In this analysis we assumed that the only background component was the particle background and detector noise.
 ACIS stowed observations  were taken for about 1 Ms. 
In this mode the detector records the particle background and detector noise.
We searched the Calibration Data Base (CALDB) for observations taken during the period
proximate to our  observations, with the same chips and  tailored ACIS background event files to 
each of our observations. 
For each pointing we re-projected the stowed observation 
to the same observed $wcs$ frame using  the CIAO tool {\em reproject\_events}; we verified that the 
stowed background events  were calibrated  with the same GAINFILE  of our observations; and we
ensured that the proper gain was used for all the ``stowed" pointings. After these procedures we extracted
the particle background spectra in the same areas described above (corresponding to CXB, uCXB and nsCXB regions).

\begin{table*}[!t]

\renewcommand{\thetable}{\arabic{table}}
\centering
\caption{Spectral analysis results} \label{tab:cts}
\begin{tabular}{cllllll}
\hline
\hline
Sample & Net-Counts  &   K$_{PL}$& $\Gamma$  &kT$_1$&kT$_2$&$\chi^2$/d.o.f \\
              & cts     &ph cm$^{-2}$ s$^{-1}$ sr$^{-1}$ & keV & keV &  &\\
\hline
\decimals
CXB &    123948   &10.91$\pm{0.16}$ ($\pm{0.26}$)  &     1.45$\pm{0.02}$ ($\pm{0.03}$) & 0.27$^{+0.02}_{-0.02}$  ($^{+0.03}_{-0.04}$) & 0.07$\pm{0.01} (\pm{0.02})$& 191.71/208  \\
uCXB  & 44642    &   4.18$\pm{0.26}$ ($\pm{0.41}$)   & 1.57$\pm{0.10}$ ($\pm{0.16}$) & 0.22$\pm{0.03}$  ($\pm{0.06})$  &  0.08$_{-0.01}^{+0.03}$ ($_{-0.03}^{+0.04}$) &  272.3/208\\
nsCXB & 11034   & 1.37$\pm{0.30}$ & 1.25$\pm{0.35}$ ($\pm{0.62}$)  & 0.22$\pm{0.04}$ ($\pm{0.05}$) &  *    & 162/105\\ 
\hline
\end{tabular}

\tablenotetext{*}{ not required}
\tablenotetext{}{ In parenthesis 90\% confidence limits.}

\end{table*}

\cite{hm06,hm07} found that the spectral shape of the {\em Chandra} background is extremely stable in time and  can be easily modeled, for extended and diffuse emission like the CXB, by using ACIS observations taken in stowed mode. 
As mentioned above, the shape of the particle background spectrum is constant in time but its amplitude is not. We
scale it by the ratio of the count rate in the [9.5-12] keV data (where no astrophysical events are recorded) 
 to that in the stowed data.  These ratios vary from 0.79-1.15. With this procedure the systematic uncertainty 
on the background estimation is $\sim$ 2\% \citep{hm06}. We averaged the background spectra of each pointing to take into account the different
locations of masked sources.
We also subtracted out of time events (counts accumulated during readouts) that account for $<$1\% of the total events. 
When using $\chi^2$ statistics, XSPEC is capable of handling systematic errors while fitting the data and adding them to the error budget. Therefore, by using the tools $grppha$, we included a 2\% systematic in the stowed background spectrum. Moreover 
\citet{lecca, hump} report that the use of $\chi^2$ or CSTAT could produce biased results in the high-counts regime. 
  According to Table  \ref{tab:cts} we expect a $<$2\% bias in the fit results. We have
  factored an additional 2\% systematic into our fits, for a total $\sim$4\% of systematics.  

While an actual risk of under-estimating the background does exist, at 1-2\% level, this risk has been mitigated by treating the flares according to their hardness ratios. Without considering the 2\% systematics, the fit does not change significantly. This is because the error budget is dominated by the intrinsic poissonian error, associated with the stowed background, which has been estimated by accumulating one third less photons than the real observation. This is clearly visible in Fig. \ref{fig:comp} where we compare the nsCXB spectrum and 
the PIB spectrum. There one can clearly note how the uncertainties are dominated by the statistical error on the PIB spectrum. 
 
Due to the uncertain background subtraction near instrumental emission lines, which  
may suffer from uncertainties of the order 5-10\% \citep{barta}, we limit our analysis to the [0.3-7] keV band and exclude 
the 2.0-2.4 keV energy range (which contains instrumental Au M$_{\alpha\beta}$ lines).

\begin{figure*}[!t]
\epsscale{1.}
\plotone{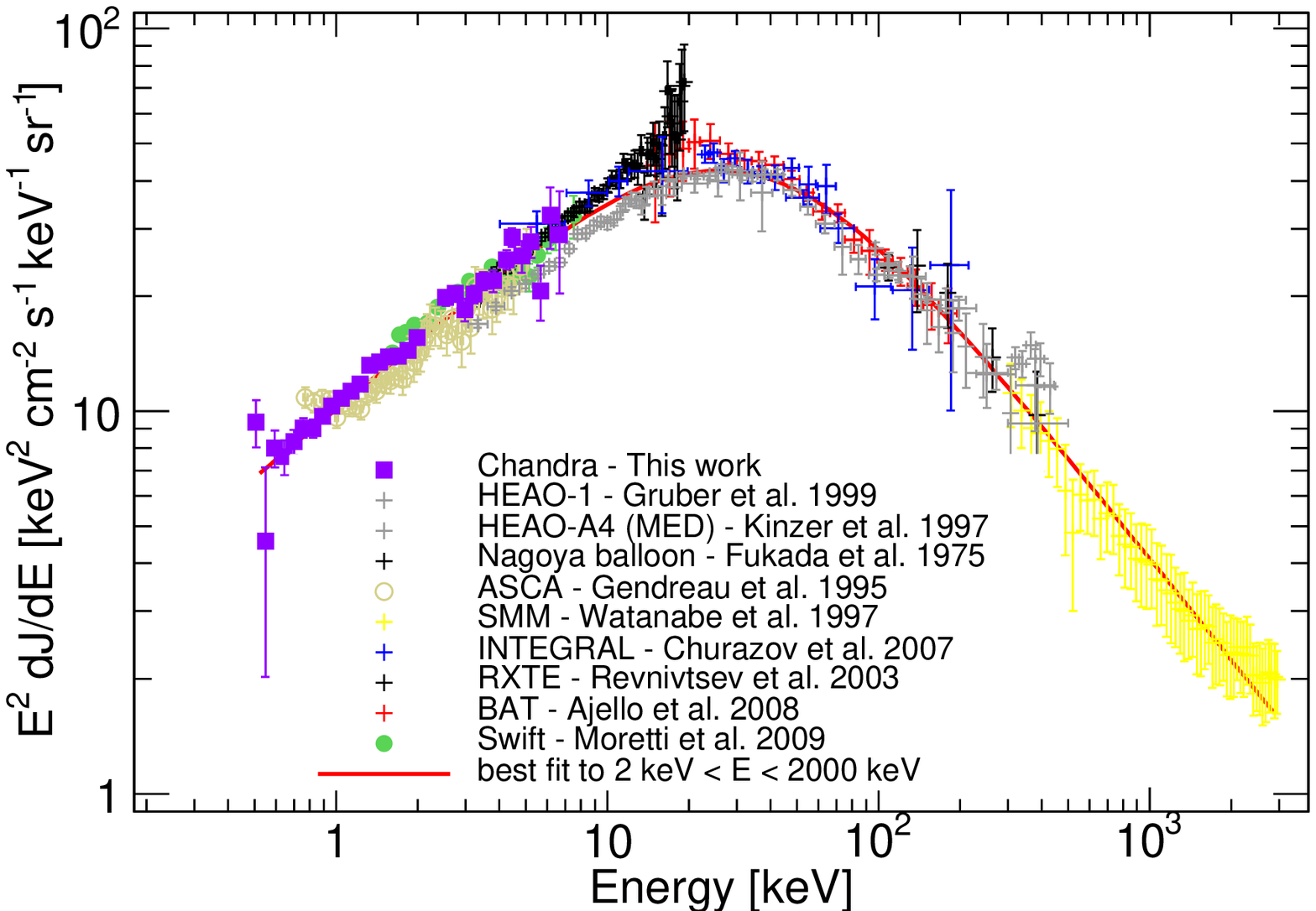}
\caption{    $Magenta~squares$  are the full CXB measured in this work using {\em Chandra} data from the COSMOS field (with local soft components subtracted),  compared with previous results over the 0.3-1000 keV energy range and the best-fit of \citet{ajello08} .
 $Green~circles$ are from \citet{more09},  $grey~crosses$ are $HEAO-1$ measurements of  
 \citet{gru} and \citet{kin}, $red~crosses$ are $Swift-BAT$ from   \citet{ajello08},  black crosses are  
 $RXTE$ from \citet{rev},  $blue~cross$ are $INTEGRAL$ from \citet{chu}, \label{fig:CXB}  $yellow~crosses$ are SMM 
 measurements from \citet{wata}, $pale~green~open~circles$ are $ASCA$ from \citet{gen}  and black~crosses $>$100 keV are from the Nagoya  balloon experiment of \citet{fuk}}
\end{figure*}

\subsection{Spectral fitting}
\begin{table*}[!t]

\renewcommand{\thetable}{\arabic{table}}
\centering
\caption{CXB fluxes} \label{tab:cxb}
\begin{tabular}{cccc}
\hline
\hline 
BAND & Total  & Local & Extragal.  \\
keV  & erg s$^{-1}$ cm$^{-2}$ deg$^{-2}$ &  erg s$^{-1}$ cm$^{-2}$ deg$^{-2}$&   \\
\hline
\hline
 0.5-1.0 &5.38$_{-0.15}^{+0.16}$    &   2.56$_{-0.18}^{+0.18}$&   3.13$_{-0.07}^{+0.07}$       \\
 1.0-2.0 &4.55$_{-0.03}^{+0.03}$  &  0.05$_{-0.01}^{+0.01}$      & 4.52$_{-0.05}^{+0.05}$ \\
 0.5-2.0 &  9.95$_{-0.18}^{+0.16}$ &  2.29  $_{-0.21}^{+0.23}$  &  7.62$_{-0.11}^{+0.11}$     \\
 2.0-10.0  &  20.34$_{-0.06}^{+0.05}$  &   0 & 20.34$_{-0.06}^{+0.05}$   \\
\hline
\hline
\end{tabular}
\tablenotetext{}{in units of 10$^{-12}$  erg s$^{-1}$ cm$^{-2}$ deg$^{-2}$ }

\end{table*}

The observed spectra are shown in Fig. \ref{fig:spectra}
We  fitted the observed X-ray spectra, grouped in bins of 
 2 channels, using XSPEC v12.9 \citep{xspec}.  
 In total the CXB and uCXB spectra have 213 spectral bins while the nsCXB has 98. 
 The full [0.3-7] keV CXB consists of 
 three principal components \citep{miyaji}: 
a) an extragalactic component produced by the integrated emission 
of resolved and unresolved discrete sources (AGN, galaxies and clusters) which we model
as a power law (hereafter PL) times Galactic absorption with N$_H$=2.0$\times$10$^{20}$ cm$^{-2}$ \citep{lock};
for the absorption we used the $tbabs$ model in XSPEC with cross-sections form \citet{verner}  and abundances from 
\citet{wilms}.
b) a Galactic hot gas component with a temperature of the order kT$\sim$0.15-0.25  keV
that we model using moderately absorbed (i.e. N$_H$=N$_{H,Gal}$) emission from 
collisionally-ionized diffuse gas (APEC, hereafter A1),  known as the hard thermal component of the
of the IGM whose temperature and intensity is a strong function of the galactic coordinates \citep[see e.g.][]{mark03}; 
c) a lower temperature local bubble and or geocoronal,  previously known as soft thermal CXB, modeled with an unabsorbed APEC  hereafter A2).  For  these CXB components we varied  the following parameters: Spectral Index $\Gamma$, PL Normalization (k$_{PL}$),  A1  and A2, normalizations and temperature kT. N$_H$ was fixed for both PL and A1. Abundances were set to solar for both A1 and A2. 

However we notice that the temperatures and amplitudes of the soft components are slightly degenerate with the power-law slope. To be conservative, we kept them free to vary in the fit.  
Moreover, on scales of several arcmin there might be fluctuations in temperature and amplitude that we want to 
take into account because of the different sizes of the field of view employed for analyzing every sub-component of the CXB.
 
 \section{Results}
 
 \begin{figure}[!h]
\epsscale{1.}
\plotone{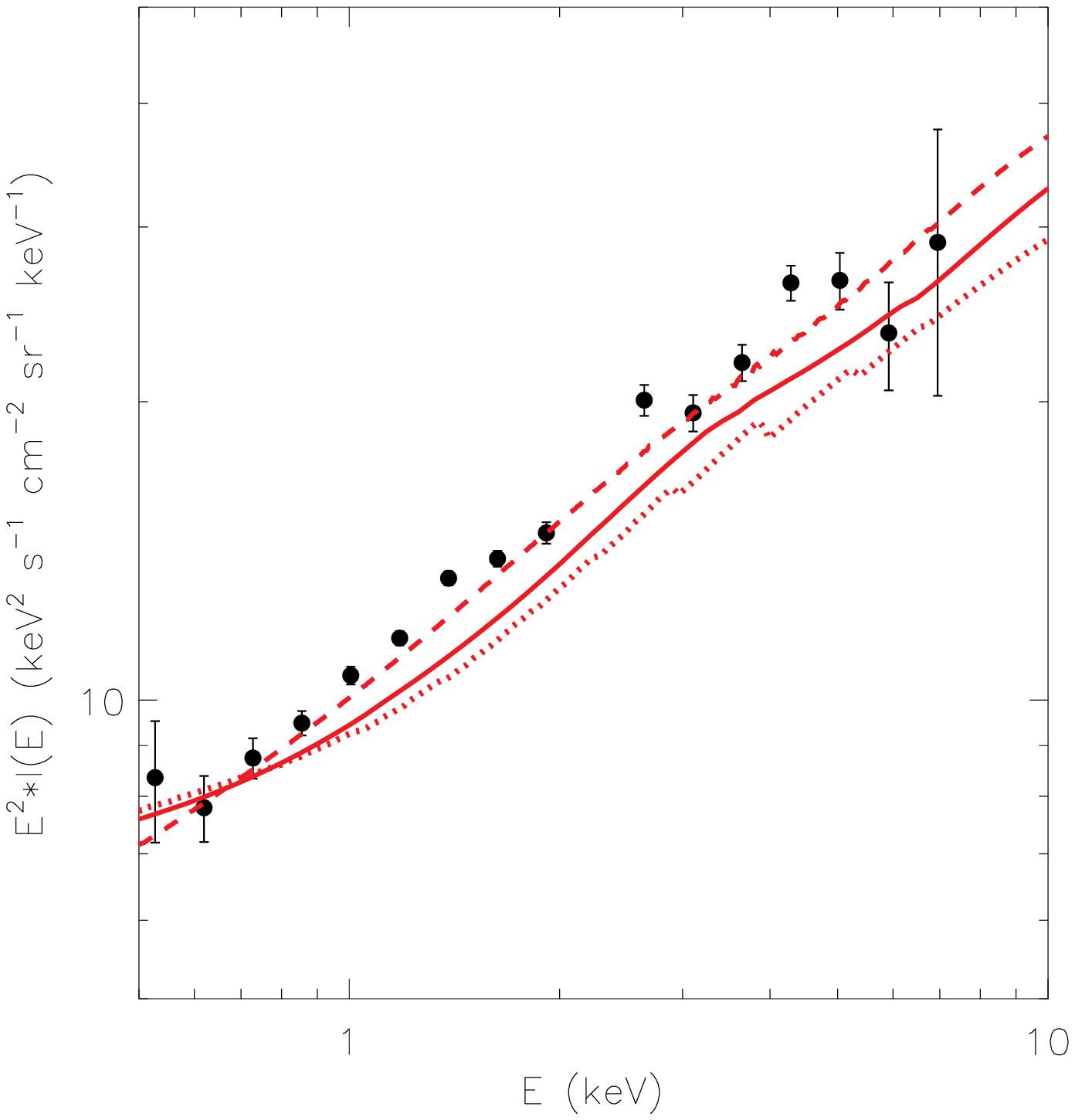}
\caption{ \label{fig:unf}  The unfolded CXB spectrum measured by    {\em Chandra}        with ACIS-I  ($black$) in this work.
Overplotted to the data we  show AGN  population synthesis models, after adding cluster emission (Gilli et al. 1999), and star 
forming galaxies emission (Cappelluti et al. 2016),  by \citet{treist09}  ($red~dotted~line$), \citet{gilli07} ($red~continuous~line$)  and \citet{balla} ($red~dashed~line$). 
The extragalactic CXB spectrum is well fitted by a power-law model with photon index Gamma$\sim$1.45 and normalization $\sim$10.91 keV cm$^2$ s$^{-1}$ sr$^{-1}$. }
\end{figure}

 \begin{figure}[!h]
\center
\epsscale{1.}
\plotone{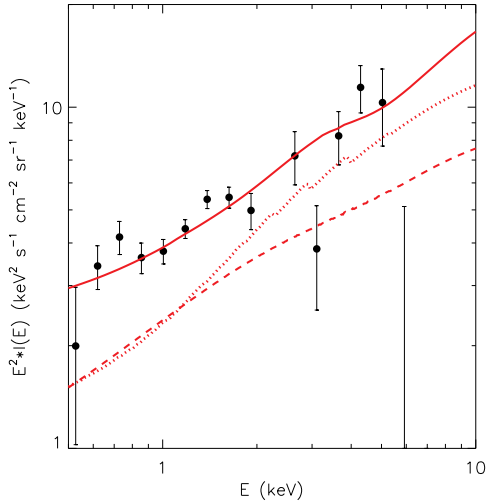}
\caption{\label{fig:comps2} The spectrum of the uCXB ($circles$) as measured with $Chandra$ ACIS-I in this work. 
The  uCXB spectrum is compared with AGN synthesis models 
 by \citet{treist09}  ($red~dotted~line$), \citet{gilli07} ($red~continuous~line$)  and \citet{balla} ($red~dashed~line$)
 after the survey X-ray selection function is applied (galactic N$_H$ 
correction is irrelevant). }
\end{figure}

 \subsection{Overall CXB spectrum}
 
 The  CXB spectrum, the best-fit model and its components are shown in Fig. \ref{fig:spectra}. 
 The best fit-parameters are summarized in Table \ref{tab:cts}. The foreground local  components have measured
 temperatures $kT\sim$0.27 keV and  $kT\sim0.07$ keV, respectively. Both the components are required at high significance level.
 Indeed, for our fit $\chi^2$/d.o.f.=191.71/208 but if we remove the local bubble (soft thermal) component we obtain  $\chi^2/d.o.f.$=231.62/208. We performed an f-test and, as a result, we obtained that the soft component is required at a 4.7$\sigma$ level. If we remove the hard thermal component the fit converges on single power-law model but with $\chi^2/d.o.f.>>$2.
   Above 2 keV, the emission can be totally ascribed to the extragalactic power-law component. The latter has a photon spectral index $\Gamma$=1.45$\pm{0.02}$,as in previous investigations  \citep[see e.g.,][]{gru,ajello08} and a normalization K$_{PL}$=10.91$\pm{0.16}$ consistent with previous    {\em Chandra}        \citep{hm06},  {\em  Swift} \citep{ajello08,more09}  and ROSAT-ASCA results  \citet{miyaji}, yet  lower than early XMM-{\em Newton} by \citet{dlm} and higher than ASCA and HEAO results \citep{gen,gru}.  This CXB  unfolded spectrum is compared with these previous  measurements in Fig. \ref{fig:CXB}. 
Due to the pencil beam nature of the survey, rare bright sources are not accounted in our measurement. A precise estimate
of their contribution is not possible with the data in hand, but using AGN population synthesis models \citep{gilli07,treist09} we estimate that our measurement of the full CXB is underestimated of 3\%. We rather  not consider this as a systematic error, but  the limitation of our total CXB measurement. 
We have also measured the overall flux of the CXB in several energy bands and reported it in Tab. \ref{tab:cxb}. The thermal contributions are dominant below 1 keV and account for about 23\% of the overall [0.5-2] keV flux. The remainder of the flux can be ascribed to extragalactic emission, while above 2 keV most of the signal is extragalactic.
The intensity of the Galactic components  is in remarkable agreement with micro-calorimeter measures of \citet{mcc}.
 
 \subsection{X-ray source  masked CXB spectrum and flux}
 
The uCXB, which is shown in Fig. \ref{fig:spectra} has a slightly softer spectrum  ($\Gamma\sim$1.57) than the overall component (CXB), but still consistent with it. The Galactic foregrounds have, within the uncertainties, the same intensity and shape as the CXB. 
The uCXB spectrum is shown in Fig. \ref{fig:comps2}. Even in this case we obtain an excellent fit with three components  with $\chi^2$/d.o.f.=272.3/208.
The slightly softer spectral slope is  consistent with the observed  higher fraction of Type I AGN among the brightest sources.
In  Table \ref{tab:ucxb} we show the fraction of resolved CXB as function of energy.  The removal of X-ray sources  
produces a drop on the remaining surface brightness of the CXB of about 70-80\% regardless of the energy. 
This is not an impressive fraction because of the CCLS flux limit, but we can compensate the shallow depth of the 
survey by searching through what is left after masking faint HST sources.
\begin{table}[!b]

\renewcommand{\thetable}{\arabic{table}}
\centering

\renewcommand{\thetable}{\arabic{table}}
\centering
\caption{Unresolved Extragalactic CXB fluxes } \label{tab:ucxb}
\begin{tabular}{ccr}
\hline
\hline 
BAND &   Extragal. &\%$_{CXB}$ \\
keV  & erg s$^{-1}$ cm$^{-2}$ deg$^{-2}$ &   \\
\hline
uCXB\\
\hline
\hlineŒ
0.5-1.0 & 1.24$\pm{0.17}$ & 23.0$\pm{3.2}$\\
1.0-2.0 & 1.66$\pm{0.06}$ &36.5$\pm{0.1}$ \\
0.5-2.0 & 2.90$\pm{0.16}$& 30.1$\pm{1.7}$  \\
2.0-10.0 & 6.47$\pm{0.82}$& 31.8$\pm{4.0}$  \\
\hline
nsCXB\\
\hline
\hline
0.5-1.0 & 0.36$_{-0.11}^{+0.13}$&6.7$_{-2.8}^{+3.0}$ \\
1.0-2.0 & 0.61$_{-0.07}^{+0.07}$& 13.4$_{-1.6}^{+1.6}$\\
0.5-2.0 &0.97$_{-0.16}^{+0.18}$ &  9.7$_{-1.8}^{+1.6}$  \\
2.0-10.0 & 3.45$_{-1.19}^{+1.42}$&   17.0$_{-7.0}^{+5.9}$ \\
\hline
\end{tabular}
\tablenotetext{}{in units of 10$^{-12}$  erg s$^{-1}$ cm$^{-2}$ deg$^{-2}$ }

\end{table}

 \subsection{Non Source CXB spectrum} 
 The extracted nsCXB cannot be directly used as is, since  
we know that up to 10\% of flux from galaxies is in the PSF tails and contaminates our results.
We have therefore extracted the spectrum of all the galaxies inside the mask, fit it with a simple absorbed power-law model (properly taking into account the different thermal background components in the fit) and found a spectral slope $\Gamma\sim$1.3$\pm{0.06}$ and normalization K=3.38$\pm{0.14}$. 
We rescaled the normalization to take into account the fraction of flux falling out
of the mask.  Such a re-normalization has been computed in the following way: 
We estimated, given a circular area of 3.2$\arcmin$, the area-weighted mean off-axis angle  $<\theta>=\frac{\int_0^{3.2^{'}} \theta*\pi\theta^2 d\theta}{\int_0^{3.2^{'}} \pi\theta^2 d\theta}$=2.25$\arcmin$. At this off-axis angle, an average of 95\% EEF is masked. Hence, we renormalized the galaxies spectrum by a factor 0.95, and simulated a spectrum taken from the rescaled best fit, accounting for all the observational parameters. We subtracted the simulated spectrum from the nsCXB 
spectrum to remove the best possible estimate of PSF tails. 
 
Due to lower statistics, to fit such a spectrum we doubled the binning with respect to cases above.  This becasue we have chosen
a binning that allowed to have at least 30 counts/bin.
In Fig. 1 we  show the spectrum of the nsCXB. The spectrum has high SNR up to an energy of 5-6 keV, above which the signal is very noisy. We allowed all parameters to vary freely, but because of the low statistics, the soft thermal component is 
detected but not significantly required. The resulting best-fit parameters are   $\Gamma$=1.25$\pm{0.35}$ and  normalization K$_{PL}$=1.37$\pm{0.30}$.  This corresponds to 9.7$^{+1.6}_{-1.8}\%$ of the total CXB in the [0.5-2] keV band. This unresolved CXB fraction is about double above 2 keV.  Our normalization of this unresolved component is in agreement with \citet{hm07} and \citet{more12}, while our estimated slope is in agreement with \citet{hm07} but significantly softer than the \citet{more12} estimate. 
In Table \ref{tab:ucxb} we show the extragalactic component flux of the nsCXB in several energy bands.  With our masking the fraction of CXB remaining varies between $\sim$6\% at very soft energies to 17\% at very high energies.

\section{Discussion}

 Ordinary 
populations of Type I and Type II AGN alone cannot explain the shape and amplitude of the extragalactic CXB spectrum, especially the peak at $\sim$30 keV \citep{com,gilli01,treist05,gilli07,treist09,balla}. Instead, this peak  is attributed to a large population of 
mostly undetected Compton-Thick sources that are naturally missed by $<10$ keV X-ray surveys.

In  Fig.  \ref{fig:unf} we compare our data with the predictions of the CXB  AGN population synthesis  models that have animated
the scientific debate in the last 10 years \citep[][]{treist05,treist09,gilli07,ueda14,balla}.  Star-forming galaxies were modeled by assuming a power-law with photon index $\Gamma$=2 and  a normalization of 0.55 keV cm$^2$ s$^{-1}$ sr$^{-1}$ keV$^{-1}$, estimated using the prescription of \citep[][see below]{cap16}. In that figure we present  the extra-galactic CXB unfolded spectrum; although such a plot is model dependent this is still a good approximation for the purpose of comparing with model. 
A cluster model component from \citet{gilli99} has been added to each of these spectra.
The three models reproduce the shape of the extragalactic CXB spectrum 
above 0.5 keV but  systematically underestimate the normalization by 10-15\%. This is likely 
due the intrinsic normalization chosen as reference for these models.
Differences among the models are of the order of the precision of our measurement.

A valuable test of the goodness of the assumptions of population synthesis models derives from whether they are 
able to reproduce the uCXB at any given flux limit. In Fig. \ref{fig:comps2} we compare the uCXB spectrum with the predictions 
of \citet{treist09,gilli07,balla}   at the flux limit of COSMOS (models assume an all sky coverage and we did not 
apply any correction for cosmic variance). Interestingly, the only model that reproduces the whole uCXB is the 
\citet{gilli07} while the \citet{treist09} is consistent with the hard X-rays. This consistency at 
high energy is not surprising since both the models aimed to explain the peak of the CXB at high
energy, and although they used different ingredients, they included a large number of hard, Compton-thick objects.   
  The  \citet{balla} model underpredicts the fraction of uCXB,
implying that their model contains more bright sources that the others. 
These discrepancies are likely due to the different assumptions for the N$_H$ distribution and  luminosity functions adopted. 
Given the quality of the CXB data we present here, the consistent  CXB levels measured by $Chandra$ and XMM-$Newton$, a new population synthesis model may be warranted.
 \begin{figure}[!ht]
\center
\epsscale{1.3}
\plotone{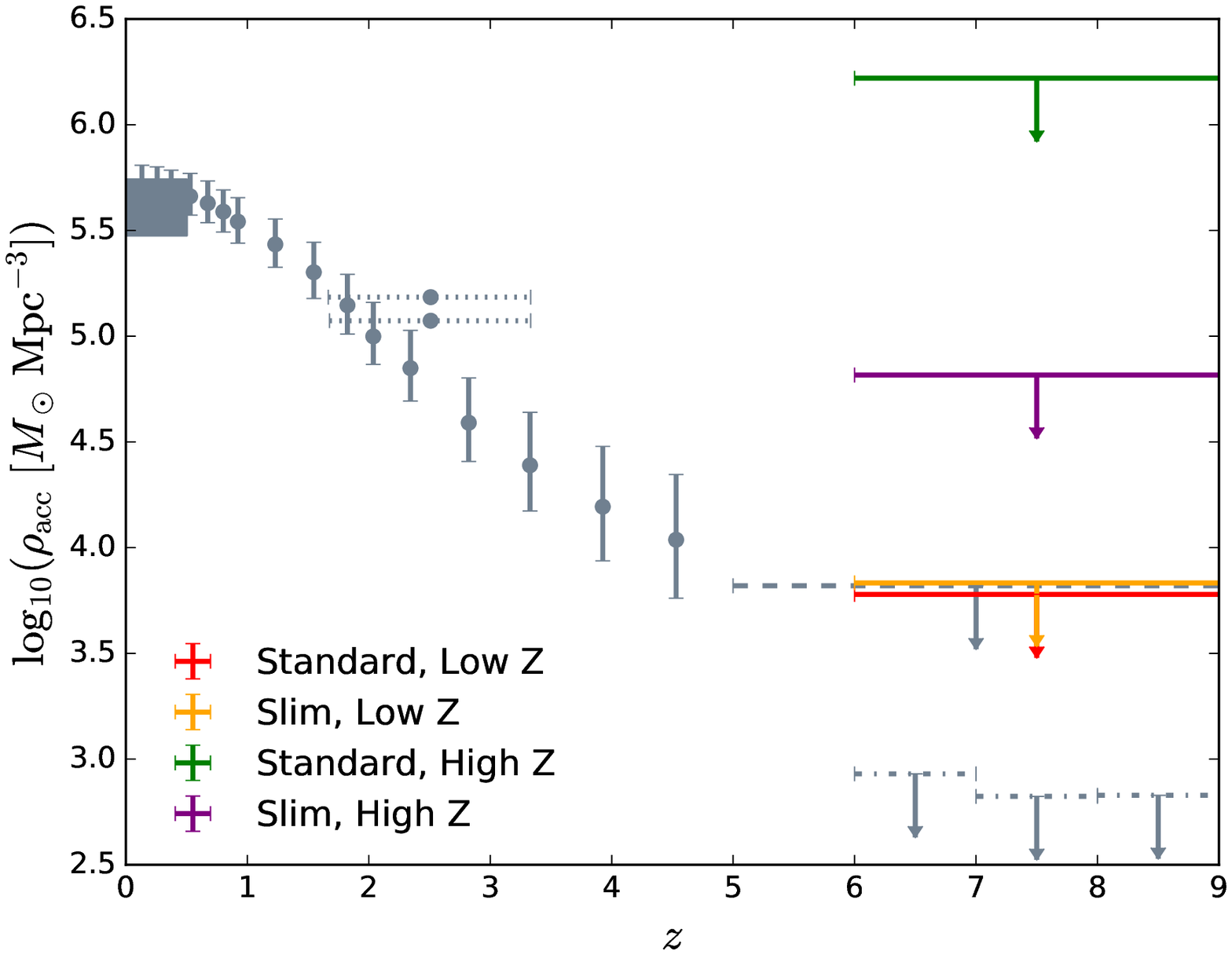}
\caption{Comparison of our limits to previous studies.  Red, orange, green, and purple bars represent our standard low-metallicity, slim disk low-metallicity, standard high-metallicity, and slim disk high-metallicity limits respectively.  In grey, solid, dashed, dotted, and dot-dashed error bars correspond to previous measurements by \citet{hop}, \citet{salva}, \citet{treist09}, \citet{treist13}.  The grey square around redshift 0 corresponds to the local estimate by \citet{sha09}.  Our study emphasizes that the assumed model dramatically changes the bolometric correction used, and therefore the limits on the accreted mass density.  \label{fig:evol}}
\end{figure}

 \begin{table*}[t!]
\footnotesize
\renewcommand{\thetable}{\arabic{table}}
\centering
\caption{Limits to the density of accretion at $z\gtrsim6$ from the unresolved background.} \label{tab:soltanArgument}
\begin{tabular}{ccc}
\hline
\hline
Accretion Disk & Metallicity $(Z_\odot)$ & $\rho_{\bullet} \,\mathrm{(M_\odot \, Mpc^{-3})}$ \\
\hline
\decimals
Standard & $10^{-3}$ & $6.1 \times 10^3$ \\
Standard & $10^{-2}$ & $1.7 \times 10^6$ \\
Slim Disk & $10^{-3}$ & $6.8 \times 10^3$ \\
Slim Disk & $10^{-2}$ & $6.5 \times 10^4$ \\

\hline
\end{tabular}
\end{table*}
According to our analysis, $\sim$8\%-11\% of the measured [0.5-2] keV CXB  cannot be explained by either resolved X-ray sources or faint, unresolved sources originating in visible red galaxies that have escaped detection.  \citet{cap12,cap13} and \citet{helga14} studied the fluctuations of 
the u/nsCXB in the deep CDFS and  EGS and concluded that these fluctuations arise from undetected groups and star forming galaxies  and a small fraction of AGN. A detailed analysis of the fluctuations of the u/nsCXB will be presented in  Li et al. (in prep). Here, we remove even fainter sources  than  \citet{cap13} and \citet{helga14},  down to i$_{AB}\sim$27-28.
Assuming that all the diffuse emission from faint groups has been 
removed by our galaxy masking,
what is left arises from very faint undetected/blurred point sources. 

In order to evaluate the contribution of star-forming-galaxies to the CXB and uCXB,  we used simulations from \citet{cap16} of the CANDELS GOODS-South area, which reaches optical/NIR magnitudes as faint as 30.  They predict for every galaxy a value of 
 L$_X$ concordant with the scaling relation with SFR (approximated by the infrared luminosity) of \citet{basu}. Without going into details,  they 
 estimate L$_{8-1000\mu m}$  using  photo-z, star formation rate, $UVJ$ rest-frame colors and  (observed or extrapolated) 
UV luminosity ($1500\AA$).  Using their mock catalog, we applied a selection as similar as possible 
to that of \citet{lai} from which  we derived our mask.
As a result we find that  these galaxies produce a [0.5-2] keV CXB surface brightness of the order 
 3.3$\times$10$^{-14}$ erg cm$^2$ s$^{-1}$ deg$^{-2}$ which explains  about 5\% of the 
 nsCXB.  By assuming a typical  X/O=0 for AGN as determined by  \citet{civ10}  and the i  limiting 
  magnitudes  in COSMOS \citep{scov}, we estimate the undetected  AGN  [0.5-2] flux is  $<$10$^{-17}$ erg cm$^2$ s$^{-1}$. 
 At these low fluxes,  star-forming-galaxies vastly outnumber AGN, so we can assume that ordinary 
AGN cannot  contribute more than galaxies to the soft nsCXB (5\%).  Being so faint, 
 the sources producing the remaining CXB can be local ($z\sim$1-3) and of low luminosity. Low luminosity AGN are preferentially highly absorbed \citep[][]{bar,has08}  and therefore we could  argue that the hard portion of the nsCXB  could be explained by these sources.  However, at z$\sim$3-6 the number density of absorbed sources is still unknown and we cannot exclude an unpredicted
  large number of such sources at that redshift.
 

We propose that a large fraction of the remaining emission could arise from still undetected, rapidly accreting, 
 black holes at z$>$6-7.
Assuming the Direct Collapse Black Hole (DCBH) scenario for the formation of  early black hole seeds, \citet{pac15} showed that these sources are likely undetected in current deep X-ray/NIR surveys. They compared the emission of DCBHs for two accretion models: radiatively efficient (Standard) and radiatively inefficient (Slim Disk; super-Eddington), in which photon trapping is significant and the outgoing radiation is diminished.
In the latter case, the luminosity emitted by these sources is  low. These short-lived and fainter black holes are more difficult to detect compared to brighter objects accreting at the Eddington limit \citep[two tentative detections were proposed by][]{pacucci}.
Indeed, \citet{com15} revised the estimate of  the local accreted mass density by taking  into account that a  significant fraction of the local black holes may have grown by radiatively inefficient accretion.
From our measurements
the maximum flux  produced by accretion onto early black holes is  $\sim$10\% of the CXB  ( see Table 3).

To place limits on the amount of accretion occurring at $z \gtrsim 6$, we follow the formalism of \citet{salva}, assuming that the comoving specific emissivity of AGN can be factorized as:
\begin{equation}
j(E,z) = j_\star f(z) g(E) \label{eqn:emissivityFactorization}
\end{equation}
where $j_\star$ is the normalization, $f(z) = (1+z)^{-\gamma}$, with $\gamma \approx 5$, is the redshift evolution, and $g(E)$ is a (normalized) template spectrum.
 For these templates, we use AGN spectra generated by realistic hydrodynamical simulations of accreting DCBHs \citep{pac15}.  These templates allow us, essentially, to compute the bolometric correction needed for the Soltan argument as a function of redshift. For each value of the gas metallicity and accretion model (Standard or Slim Disk), we select the spectrum from the snapshot with the highest X-ray output. 
 Combining equation \ref{eqn:emissivityFactorization} with knowledge of the contribution to the background at energy $E_0$ by sources at redshifts $z \geq \bar{z}$, the normalization can be solved for:
\begin{equation}
\begin{split}
j_\star = \frac{4\pi J_{E_0} H_0 \Omega_m^{1/2}}{c} \left[ \int_{\bar{z}}^{\infty} dz (1+z)^{-5/2-\gamma} g(E_0(1+z)) \right]^{-1} \label{eqn:normalization}
\end{split}
\end{equation}
where   J$_{E_0}$ is the emissivity observed at energy E$_0$ today,  j(E,z) on the other hand is the emissivity of all AGN at redshift z. Note that E is in the rest frame, and E$_0$ is the energy observed at z=0. $\Omega_m$ is matter density parameter and H$_0$ is the Hubble constant.

The standard Soltan argument states that the mass density of accretion onto sources at redshifts $z \geq \bar{z}$ is given by:
\begin{equation}
\begin{split}
\rho_\mathrm{acc}(\bar{z}) = \frac{1-\epsilon}{\epsilon c^2} \int_{\bar{z}}^{\infty} dz \frac{dt}{dz} \int_{0}^{\infty} dE j(E,z) \label{eqn:Soltan}
\end{split}
\end{equation}
where $\epsilon$ is the radiative efficiency ($0.1$ and $\lesssim 0.04$ for a standard and a slim disk, respectively.  Finally, our limit on accretion at $z \geq \bar{z}$ is given by
\begin{equation}
\begin{split}
\rho_\mathrm{acc}(\bar{z}) = \frac{4 \pi J_{E_0}}{c^3} \frac{1-\epsilon}{\epsilon} \left[\int_{\bar{z}}^{\infty} dz (1+z)^{-5/2-\gamma} \right] \\\left[ \int_{\bar{z}}^{\infty} dz (1+z)^{-5/2-\gamma} g(E_0(1+z)) \right]^{-1} .\label{eqn:accretionLimit}
\end{split}
\end{equation}
Assuming that the unresolved 1.5 keV flux is entirely due to DCBHs  at $z \gtrsim 6$, the inferred accretion density ($\rho_{\bullet}$)  using these spectra is provided in Table \ref{tab:soltanArgument}.  These limits are compared to those found in previous studies in Fig. \ref{fig:evol}.  Red, orange, green, and purple upper limits correspond to standard low-metallicity, slim disk low-metallicity, standard high-metallicity, and slim disk high-metallicity templates respectively.  In grey, we display limits from previous studies.  The straight, dashed, dotted, and dot-dashed error bars correspond to measurements from \citet{hop}, \citet{salva}, \citet{treist09}, and \citet{treist13} respectively.  The grey square corresponds to local measurements by \citet{sha09}.  Our results emphasize that limits to black hole accretion are dependent entirely on the bolometric correction assumed, and this can vary significantly from model to model.  Again, while previous studies have assumed a constant fraction of total flux emitted in the observed window, we calculated this fraction directly from hydrodynamical simulations.  These models imply that much less accretion is required to provide the observed flux if gas is accreted from a lower-metallicity reservoir $Z_\odot=10^{-3}$ while larger metallicities  $Z_\odot>10^{-2}$ would exceed the z$\sim$5-6 accreted density of \citet{hop}.   The limits in the lower metallicity case are comparable to or more stringent than what is obtained with stacking analysis in \citet{treist13} ($\rho_{\bullet} \lesssim 10^{3} \, \mathrm{M_{\odot} \, Mpc^{-3}}$). 
 
  \citet{cap13} determined that the unresolved nsCXB and unresolved cosmic infrared background  fluctuations are highly correlated \citep[see e.g.][]{kamm12}.  \citet{yue13} interpreted this as signature of emission from a 
population of DCBHs  at z$>$12. In order to satisfy the observed cross-power and not to 
exceed the nsCXB measured here, their envelopes must be Compton-thick. With our new limits on the nsCXB, according to \citet{yue13}  DCBHs must have N$_H>$1.6$\times$10$^{25}$ cm$^{-2}$. To summarize if this population of early massive black holes exist, they had to grow in Compton-thick, low-metallicity environments.

 \acknowledgments
 NC acknowledges the Yale University's YCAA Prize Postdoctoral fellowship. NC, GH, YL
and FP acknowledge the SAO {\em Chandra} grant AR6-17017B. and NASA-ADAP grant MA160009.
PN acknowledges support from a Theoretical and Computational Astrophysics Network grant with award number
1332858 from the National Science Foundation. BA and AR acknowledge support from
the TCAN grant for a post-doctoral fellowship and a graduate fellowship respectively.
AC e RG acknowledge PRIN INAF 2014 Ð Ò Windy black holes combing galaxy evolutionÓ 
and ASI/INAF grant I/037/12/0Ð 011/13. ET acknowledge support from FONDECYT regular grant 1160999 and Basal-CATA PFB-06.
We thank the anonymous referee for the useful insights and suggestions.

\end{document}